\documentclass[12pt]{article}

\newcommand{\bm}[1]{\mbox{\boldmath $#1$}}
\def\espaitemps{({\cal V},g)}

\def\S{\Sigma}

\def\be{\begin{equation}}
\def\ee{\end{equation}}
\def\bea{\begin{eqnarray}}
\def\eea{\end{eqnarray}}
\def\bean{\begin{eqnarray*}}
\def\eean{\end{eqnarray*}}

\begin{document}

\title{Symmetric hyperbolic systems for a large class of fields in arbitrary dimension}
\author{Jos\'e M.M. Senovilla \\
F\'{\i}sica Te\'orica, Universidad del Pa\'{\i}s Vasco, \\
Apartado 644, 48080 Bilbao, Spain \\ 
josemm.senovilla@ehu.es}
\date{}
\maketitle
\begin{abstract} 
Symmetric hyperbolic systems of equations are explicitly constructed for a general class of tensor fields by considering their structure as $r$-fold forms. The hyperbolizations depend on $2r-1$ arbitrary timelike vectors. The importance of the so-called ``superenergy" tensors, which provide the necessary symmetric positive matrices, is emphasized and made explicit. Thereby, a unified treatment of many physical systems is achieved, as well as of the sometimes called ``higher order" systems. The characteristics of these symmetric hyperbolic systems are always physical, and directly related to the null directions of the superenergy tensor, which are in particular principal null directions of the tensor field solutions. Generic energy estimates and inequalities are presented too.
\end{abstract} 


\section{Introduction}
\label{intro}
First order symmetric hyperbolic systems of partial differential equations are of paramount importance in mathematical physics. Since the pioneering work by Friedrichs \cite{Fri}, many such systems have been studied in different areas. In gravitational physics, they are relevant both in studying particular physical systems on given spacetime backgrounds, and in considering the gravitational field equations themselves ---be them Einstein's field equations or more general possibilities---. Their importance resides on the availability of very powerful theorems of existence and uniqueness of solutions, under mild continuity or differentiability assumptions.

Hyperbolicity of a system of partial differential equations is a concept which generally implies the well-posedness of the appropriate Cauchy problem, that is to say, a well-defined initial value formulation. The general reason behind this is the existence of norms ---on the space of solutions--- which are well-behaved under the evolution defined by the system. One of the purposes of this short paper is to bring out the relevance that the so-called `superenergy' tensors have in the definition of these norms and on the hyperbolizations of general systems in Lorentzian backgrounds of arbitrary dimension.

Symmetric hyperbolic systems for most of the known physical fields can be set up, in both flat and non-flat spacetimes, with more or less difficulty. For an account of these, one can consult  \cite{G,R,Be} and references therein. In these cases, the background spacetime is a given Lorentzian manifold and the field equations can be written with the metric, connection and curvature variables as known data. A much more difficult problem is that of solving the field equations for a physical field coupled to gravity, for the background spacetime (metric, connection, etc.) must be built at the same time by solving the corresponding field equations. In this case, the causal structure inherent in the symmetric hyperbolic systems and that corresponding to the spacetime thus constructed should agree, leading to causal propagation of gravity and physical fields. The importance of superenergy tensors in the study of causal propagation was already made explicit in \cite{BS,BoS}.

In recent years there has been an increase of attention concerning symmetric hyperbolic systems for Einstein's field equations due to the exigencies of the developing area known as Numerical Relativity, see e.g. \cite{BMSS,FR,Re,S} and references therein, where initial data are evolved by numerical integration of the appropriate field equations. 
As a matter of fact, the history of symmetric hyperbolic systems in General Relativity is largely related to the different developments concerning the Cauchy problem for the Einstein field equations, see \cite{ChY,FM,FR} and references therein. For very good recent surveys of the gravitational Cauchy problem, see \cite{KN,FR}. A new input to this problem came with the work by Friedrich \cite{F,F1} where the Bianchi identities for the Weyl curvature tensor were included in the system, and the Bel-Robinson `superenergy' tensor was utilized to estimate its strength. This was later improved in \cite{ChY1}, where only physical characteristics were obtained and matter sources could be included. These results relied on an `electric-magnetic' decomposition of the Riemann tensor, and on its derived ``superenergy", both introduced by Bel \cite{XYZZ} in 4 dimensions many years ago. Improved results of this approach were found in \cite{AChY}, where an integral inequality for the Bel superenergy was introduced and causal propagation of the Riemann tensor in vacuum was also obtained. In these and later papers \cite{ChY3,ChY2}, all the results were derived in four dimensions. 

An interesting approach to these results was given in \cite{Bo}, by just presenting direct hyperbolizations ---in the sense of \cite{G}--- of the Bianchi equations. Again the results were obtained in 4 dimensions. However, the whole argument in \cite{Bo} is valid, {\it mutatis mutandis}, in arbitrary dimensions {\it except} for the crucial final step where the positivity properties of the system's symmetric matrix was proved. To that end, in \cite{Bo} spinors were used\footnote{Actually, by using spinors the hyperbolizations of the Bianchi identities, and of quite general spinor fields, are rather obvious, see \cite{G,F1,FR}}, which restricts the result to 4 dimensions exclusively. 

This is a key point in the present paper: for some time now it has been known that the electric-magnetic decomposition of the Riemann tensor, and actually of {\em arbitrary} tensors, as well as the superenergy construction and the positivity properties of the superenergy tensors, {\em hold in arbitrary dimensional} Lorentzian manifolds and furthermore, universally, that is to say, for arbitrary tensor fields. The basic references for this are \cite{S0,S} and references therein. 

By using those facts, I am going to show how to construct, in general Lorentzian manifolds of arbitrary dimension, first-order symmetric hyperbolic systems of equations for general tensor fields subject to suitable field equations, and the crucial role that the positivity properties of the superenergy tensors play. Let $\espaitemps$ be an $n$-dimensional Lorentzian manifold with metric tensor $g$ and signature $(-,+,\dots,+)$. The archetypical system to be hyperbolized is
\be
\nabla_{[\mu_0}A_{\mu_1\dots\mu_s]\mu_{s+1}\dots \mu_m}=J_{\mu_0\dots \mu_m}, \hspace{1cm}
\nabla^{\rho}A_{[\rho\mu_2\dots\mu_s]\mu_{s+1}\dots \mu_m}=j_{\mu_2\dots \mu_m}\label{paradigm}
\ee
where the unknown $A_{\mu_1\dots \mu_m}$ is an {\em arbitrary} rank-$m$ tensor field, 
$J_{\mu_0\dots \mu_m}=J_{[\mu_0\dots \mu_s]\mu_{s+1}\dots \mu_m}$ and 
$j_{\mu_2\dots \mu_m}=j_{[\mu_2\dots\mu_s]\mu_{s+1}\dots \mu_m}$ are given and may depend on the background Lorentzian manifold and on the tensor field $A_{\mu_1\dots \mu_m}$ (but not on its first derivatives), and $s$ is any natural number such that $0\leq s\leq \min\{m,n\}$. This is not the only type of system susceptible of study by superenergy techniques, but it will serve to make the main points. Other more general systems can also be treated in the same manner, and some comments and examples will be made in section \ref{apps}. In particular, one can deal with several, possibly interacting, fields by either (i) letting $A_{\mu_1\dots \mu_m}$ to be an inhomogeneous `multi-tensor' as in \cite{PP} or (ii) by adding systems similar to (\ref{paradigm}) for the other fields. The latter possibility will be adopted here, see subsection \ref{mixed}. The Bianchi equations with and without sources, in arbitrary dimension, will be dealt with in subsection \ref{Bianchi}.

The approach presented herein has the virtues of (i) unifying many different systems of equations, (ii) opening new possible applications for yet unexplored cases, (iii) providing a simple interpretation of the characteristics of the systems, (iv) providing in a direct general manner integral inequalities for the ``energy" of the systems (which is nothing but the ``superenegry density" \cite{S}), (v) allowing us to deal with systems with only one of the two expressions in (\ref{paradigm}) by adding `gauge' equations, and (vi) giving, under some circumstances, natural divergence-free tensor fields as well as conserved quantities.

\section{The `superenergy' construction}
\label{s-e}
First of all, some fundamental properties of the superenergy construction and the superenergy tensors need to be recalled. Everything is based on the following basic result \cite{S0,S}, see also \cite{ES}: any tensor $A_{\mu_1\dots \mu_m}$ can be considered, in a precise and unique way, as an {\em $r$-fold form}, that is to say, as a tensor belonging to $\Lambda^{n_{1}}\otimes\dots\otimes \Lambda^{n_{r}}$ where $\Lambda^p$ is the set of $p$-forms.  
The number $r$ is a well defined natural number called the {\em form-structure number} of $A_{\mu_1\dots \mu_m}$ and, obviously, $r\leq m$. Similarly, the set of $r$ natural numbers $n_{1},\dots ,n_{r}$ is {\em uniquely} defined, and each $n_\Upsilon$ is called the {\em $\Upsilon$-th block rank}. It is trivial that  $\sum_{\Upsilon =1}^{r}n_{\Upsilon}=m$, and $n_\Upsilon\leq \min\{n,m\}$ for all $\Upsilon = 1,\dots ,r$ hence $m/n\leq r\leq m$. Tensors seen in this way are called $r$-fold $(n_{1},\dots ,n_{r})$-forms and denoted by $A_{[n_1]\dots[n_r]}$ \cite{S}. A tilde on a tensor ($\tilde{A}_{\mu_1\dots \mu_m}$, or equivalently, $\tilde{A}_{[n_1]\dots[n_r]}$) indicates that the indices have been permutted so that the first $n_1$ indices of $\tilde{A}_{\mu_1\dots \mu_m}$ are those precisely in $[n_1]$, the next $n_2$ indices are those in $[n_2]$, and so on.\footnote{Some simple examples are: any $p$-form $\S_{\mu_1\dots\mu_p}=\S_{[\mu_1\dots\mu_p]}$ is trivially a 
single (that is, 1-fold) $p$-form, while $\nabla_\nu\S_{\mu_1\dots\mu_p}$ is a double 
$(1,p)$-form. The Riemann tensor $R_{\alpha\beta\lambda\mu}$ is a double {\em symmetric} (2,2)-form and the Ricci tensor $R_{\beta\mu}$ is a double symmetric (1,1)-form. In general, any completely symmetric 
rank-$m$ tensor is an $m$-fold (1,1,\dots ,1)-form. A rank-3 tensor $A_{\alpha\beta\gamma}$ with the 
property $A_{\alpha\beta\gamma}=-A_{\gamma\beta\alpha}$ is a double 
(2,1)-form and the corresponding $\tilde A$ is clearly given by 
$\tilde{A}_{\alpha\beta\gamma}=\tilde{A}_{[\alpha\beta]\gamma}\equiv A_{\alpha\gamma\beta}$.} 

Once the indices are split into skew-symmetric blocks, the canonical `Electric-Magnetic' (E-H in short) decomposition of the tensor, associated with {\em any} unit timelike vector $\vec u$, is immediately obtained. Probably the most natural way to do this is by using the Hodge dual operators $*_{\Upsilon}$ acting on each block \cite{S0,S,ES}. Thus, there are exactly $2^r$ electric-magnetic parts of a given tensor whose form structure number is $r$.\footnote{Of course, if the original tensor has extra symmetries, or if some of its traces vanish, then there may be some relations between different E-H parts. Thus, for instance, the two mixed E-H parts of the Riemann tensor are related, and there are some extra properties for the Weyl tensor $C_{\alpha\beta\lambda\mu}$. The fact that in $n=4$ there are only one electric and one magnetic part of $C_{\alpha\beta\lambda\mu}$ is a purely {\em dimensional} result; for further results and explanations, see \cite{S,S1}.} For each block, the contraction with $\vec u$ produces the electric part in that block, and the contraction with the dualized block the corresponding magnetic part. Another way of seeing this is by contracting each block with $\vec u$ (electric) and then taking the wedge or exterior product of that block with $\vec u$ (magnetic). Yet another even simpler way of thinking of this is by taking, in any orthonormal basis $\{\vec{e}_0,\vec{e}_1,\dots ,\vec{e}_{n-1}\}$ with $\vec u=\vec{e}_0$, the $0i_2\dots i_{n_\Upsilon}$-components  (electric) and the $i_1\dots i_{n_\Upsilon}$-components (magnetic) in the selected block, where Latin small indices take the values $1,\dots ,n-1$. All the E-H parts are spatial tensors (orthogonal to $\vec u$) and they completely characterize the tensor field.

Given any tensor $A_{\mu_1\dots\mu_m}$ with form structure number $r$, its basic superenergy tensor $T_{\lambda_1\mu_1\dots\lambda_r\mu_r}\{A\}$ has $2r$ indices, and it is symmetric on each of the $r$ pairs \cite{S}. The standard definition of the basic superenergy tensor is given with Hodge duals, but a very useful property is that they can be written, actually, independently of the dimension $n$ and without any duals, see section 3 in \cite{S}. This implies that one can actually define a `superenergy operator', valid in arbitrary dimension $n$, whose outcome when applied to the tensor product $A\otimes A$ is the basic superenergy tensor of $A$. This operator reads
\bea
E_{\lambda_1\mu_1\dots\lambda_r\mu_r}
{}^{\sigma_1\dots\sigma_{n_1}\dots\tau_1\dots\tau_{n_r}}_{\rho_1\dots\rho_{n_1}\dots \nu_1\dots\nu_{n_r}}\equiv \frac{1}{(n_1-1)!} \delta^{\sigma_2\dots\sigma_{n_1}}_{\rho_2\dots\rho_{n_1}}\left(2\delta^{\sigma_1}_{(\lambda_1}g_{\mu_1)\rho_1}-\frac{1}{n_1}\delta^{\sigma_1}_{\rho_1}g_{\lambda_1\mu_1}\right)\times \cdots \nonumber\\
\times \frac{1}{(n_r-1)!} \delta^{\tau_2\dots\tau_{n_r}}_{\nu_2\dots\nu_{n_r}}\left(2\delta^{\tau_1}_{(\lambda_r}g_{\mu_r)\nu_1}-\frac{1}{n_r}\delta^{\tau_1}_{\nu_1}g_{\lambda_r\mu_r}\right)\label{E}
\eea
where
$$
\delta^{\mu_1\dots\mu_p}_{\nu_1\dots\nu_p}\equiv p! \delta^{\mu_1}_{[\nu_1}\cdots\delta^{\mu_p}_{\nu_p]}
$$
is the Kronecker symbol of order $p$. In other words, the operator consists of a factor
$$
\frac{1}{(n_\Upsilon-1)!} \delta^{\sigma_2\dots\sigma_{n_\Upsilon}}_{\rho_2\dots\rho_{n_\Upsilon}}\left(2\delta^{\sigma_1}_{(\lambda_\Upsilon}g_{\mu_\Upsilon)\rho_1}-\frac{1}{n_\Upsilon}\delta^{\sigma_1}_{\rho_1}g_{\lambda_\Upsilon\mu_\Upsilon}\right)
$$
for each antisymmetric block $[n_\Upsilon]$ of the tensor $A_{[n_1]\dots[n_r]}$. The basic superenergy tensor of $A_{\mu_1\dots\mu_m}$ is then simply
\be
T_{\lambda_1\mu_1\dots\lambda_r\mu_r}\{A\}=\frac{1}{2} 
E_{\lambda_1\mu_1\dots\lambda_r\mu_r}
{}^{\sigma_1\dots\sigma_{n_1}\dots\tau_1\dots\tau_{n_r}}_{\rho_1\dots\rho_{n_1}\dots \nu_1\dots\nu_{n_r}}\, \tilde{A}_{\sigma_1\dots\sigma_{n_1}\dots\tau_1\dots\tau_{n_r}}\tilde{A}^{\rho_1\dots\rho_{n_1}\dots\nu_1\dots\nu_{n_r}} . \label{set}
\ee
Observe, as remarked above, that this is independent of the dimension $n$. Note also that once this formula is known, one may let the operator act on two {\em different} tensors $A_{\mu_1\dots\mu_m}$ and $B_{\mu_1\dots\mu_m}$ as long as they have the {\em same form structure number and the same block ranks}, thus defining an operator acting on the tensor product $A\otimes B$:
$$
T_{\lambda_1\mu_1\dots\lambda_r\mu_r}\{A,B\}=\frac{1}{2} E_{\lambda_1\mu_1\dots\lambda_r\mu_r}
{}^{\sigma_1\dots\sigma_{n_1}\dots\tau_1\dots\tau_{n_r}}_{\rho_1\dots\rho_{n_1}\dots \nu_1\dots\nu_{n_r}}\, \tilde{A}_{\sigma_1\dots\sigma_{n_1}\dots\tau_1\dots\tau_{n_r}}\tilde{B}^{\rho_1\dots\rho_{n_1}\dots\nu_1\dots\nu_{n_r}} .
$$
An equivalent way of defining this operation is (suppressing indices)
$$
T\{A,B\}\equiv 2\, T\left\{\frac{A+B}{2}\right\}-\frac{1}{2}T\{A\}-\frac{1}{2}T\{B\}
$$
so that $T\{A,B\}=T\{B,A\}$ and $T\{A,A\}=T\{A\}$.

\section{Hyperbolizations}
\label{hyper}
There are three inequivalent possibilities for the system (\ref{paradigm}) according to whether $s$ is lower, equal, or greater than the first block rank $n_1$ of $A$. (Of course, one could also study the system (\ref{paradigm}) by antisymmetrizing over any chosen $s$ indices of the tensor field, and not necessarily those including the first one. The argument will work just the same, so that this must be understood as a way of making the reasoning clearer without affecting the generality.) If $s<n_1$ (including the case $s=0$), then in fact the system is equivalent to
\be
\nabla_{\mu_0}A_{\mu_1\dots \mu_m}=(s+1)J_{\mu_0[\mu_1\dots\mu_{n_1}]\mu_{n_1+1}\dots\mu_m}-(-1)^{n_1}sJ_{[\mu_1\dots\mu_{n_1}]\mu_0\mu_{n_1+1}\dots\mu_m},\label{trivial}
\ee
so that the divergence equation $\nabla^{\rho}A_{\rho\mu_2\dots \mu_m}=j_{\mu_2\dots \mu_m}$
follows from (\ref{trivial}) and thus 
$$
j_{\mu_2\dots \mu_m}=(s+1)J^{\rho}{}_{[\rho\mu_2\dots\mu_{n_1}]\mu_{n_1+1}\dots\mu_m}-(-1)^{n_1}sJ_{[\rho\mu_2\dots\mu_{n_1}]}{}^{\rho}{}_{\mu_{n_1+1}\dots\mu_m}
$$  
for consistency. Equations (\ref{trivial}) are very simple and easily hyperbolized ---see footnote \ref{foot2} below---.
If $s>n_1$ then there are equations {\em only} for the new tensor field $\tilde{A}'_{\mu_1\dots\mu_m}\equiv \tilde{A}_{[\mu_1\dots\mu_s]\mu_{s+1}\dots \mu_m}$ which now has $s=n'_1$. Finally, the most interesting case is precisely when $s=n_1$, which includes the previous one.\footnote{The system (\ref{paradigm}) for $s=n_1$ can be written in an intrinsic manner by using the exterior differential and co-differential acting on the blocks of $r$-fold forms, see \cite{ES} and references therein. With the notation introduced in \cite{ES}, the system reads simply $d_{(1)}{A}=(s+1){J}$ and $\delta_{(1)} {A}=-{j}$.\label{foot}}

In order to find the hyperbolizations of (\ref{paradigm}) in this case, recall a very important property of the basic superenergy tensors:
the tensor (\ref{set}) is the essentially unique (i.e., unique up to index permutations; see section 5 of \cite{S}) tensor quadratic on $A_{\mu_1\dots\mu_m}$ which satisfies the {\em dominant property} \cite{S,BS1}. A convenient way of stating this fundamental property for the purposes of this paper is that, for any set of arbitrary timelike future-directed vectors $\{u_1^{\lambda_1},v_1^{\mu_1},\dots,u_r^{\lambda_r},v_r^{\mu_r}\}$,
\be
T_{\lambda_1\mu_1\dots\lambda_r\mu_r}\{A\}u_1^{\lambda_1}v_1^{\mu_1}\dots u_r^{\lambda_r}v_r^{\mu_r}>0\label{DP}
\ee
(for non-zero $A_{\mu_1\dots\mu_m}$). An equivalent statement is that the vectors
$$
P^{\alpha}=-T^{\alpha}{}_{\mu_1\dots\lambda_r\mu_r}\{A\}v_1^{\mu_1}\dots u_r^{\lambda_r}v_r^{\mu_r}
$$
are causal and future directed. It follows that the vector-valued matrices (endomorphisms acting on the set of $r$-fold $(n_1,\dots,n_r)$-forms) defined, for arbitrary timelike future-directed vectors $\{v^{\mu_1},u_2^{\lambda_2},v_2^{\mu_2},\dots,u_r^{\lambda_r},v_r^{\mu_r}\}$, by
\be
Q^{\alpha}\,{}{}^{\sigma_1\dots\sigma_{n_1}\dots\tau_1\dots\tau_{n_r}}_{\rho_1\dots\rho_{n_1}\dots\nu_1\dots\nu_{n_r}}\equiv 
E^{\alpha}{}_{\mu_1\lambda_2\mu_2\dots\lambda_r\mu_r}
{}^{[\sigma_1\dots\sigma_{n_1}]\dots[\tau_1\dots\tau_{n_r}]}_{[\rho_1\dots\rho_{n_1}]\dots[\nu_1\dots\nu_{n_r}]}
v^{\mu_1}u_2^{\lambda_2}v_2^{\mu_2}\dots u_r^{\lambda_r}v_r^{\mu_r}\label{Q}
\ee
are appropriate candidates for the symmetric positive-definite matrix of a symmetric hyperbolic form of (\ref{paradigm}). Indeed, for arbitrary $r$-fold $(n_1,\dots,n_r)$-forms $A_{[n_1]\dots[n_r]}$ and $B_{[n_1]\dots[n_r]}$ one has that
\bean
Q^{\alpha}(A,B)&\equiv& Q^{\alpha}\,{}{}^{\sigma_1\dots\sigma_{n_1}\dots\tau_1\dots\tau_{n_r}}_{\rho_1\dots\rho_{n_1}\dots\nu_1\dots\nu_{n_r}}
\tilde{A}_{\sigma_1\dots\sigma_{n_1}\dots\tau_1\dots\tau_{n_r}}
\tilde{B}^{\rho_1\dots\rho_{n_1}\dots\nu_1\dots\nu_{n_r}}=\\
&=&T^{\alpha}{}_{\mu_1\lambda_2\mu_2\dots\lambda_r\mu_r}\{A,B\}
v^{\mu_1}u_2^{\lambda_2}v_2^{\mu_2}\dots u_r^{\lambda_r}v_r^{\mu_r}
\eean
so that on the one hand
$$
Q^{\alpha}(A,B)=Q^{\alpha}(B,A)\, ,
$$
and on the other hand, for any timelike future-directed 1-form $u_{\alpha}$, $u_{\alpha}Q^{\alpha}(\cdot,\cdot)$ is positive definite as follows from ($v^{\mu}=v_1^{\mu}$)
$$
u_{\alpha}Q^{\alpha}(A,A)=-u_{\alpha}P^{\alpha}>0\, .
$$
Now it is easy to find hyperbolizations (in the sense of \cite{G}) of the system (\ref{paradigm}) for $s=n_1$. First, rewrite (\ref{paradigm}) as 
\bea
\left(-\frac{1}{(s-1)!} g^{\alpha[\sigma_1}\delta^{\sigma_2\dots\sigma_{s}]}_{\mu_2\dots\mu_s},
\frac{1}{(s+1)!}\delta^{\alpha\sigma_1\dots\sigma_s}_{\mu_0\mu_1\dots\mu_s}\right)\frac{1}{n_2!}\delta^{\rho_1\dots\rho_{n_2}}_{\beta_1\dots\beta_{n_2}}\cdots\frac{1}{n_r!}\delta^{\tau_1\dots\tau_{n_r}}_{\nu_1\dots\nu_{n_r}}\nabla_{\alpha}\tilde{A}_{\sigma_1\dots\sigma_s\rho_1\dots\rho_{n_2}\dots\tau_1\dots\tau_{n_r}}\nonumber\\
=\left(j_{\mu_2\dots\mu_s\beta_1\dots\beta_{n_2}\dots\nu_1\dots\nu_{n_r}}, 
J_{\mu_0\mu_1\dots\mu_s\beta_1\dots\beta_{n_2}\dots\nu_1\dots\nu_{n_r}}\right)\hspace{3cm}
\label{para2}
\eea
and now contract  this with
\bea
\frac{1}{2(s-1)!}
\left(-\frac{1}{(s-1)!} v_{[\gamma_1}\delta^{\mu_2\dots\mu_s}_{\gamma_2\dots\gamma_s]}\, ,
-\frac{s+1}{s!\, s}\, v^{[\mu_0}\delta^{\mu_1\dots\mu_s]}_{\gamma_1\dots\gamma_s}\right)\times\nonumber\\
\frac{1}{(n_2-1)!}
\left(u_2^{[\beta_1}v_{2[\epsilon_1}\delta^{\beta_2\dots\beta_{n_2}]}_{\epsilon_2\dots\epsilon_{n_2}]}+
v_2^{[\beta_1}u_{2[\epsilon_1}\delta^{\beta_2\dots\beta_{n_2}]}_{\epsilon_2\dots\epsilon_{n_2}]}-
\frac{u_2^{\rho}v_{2\rho}}{n_2}\delta^{\beta_1\dots\beta_{n_2}}_{\epsilon_1\dots\epsilon_{n_2}}\right)
\times \cdots \label{hyp}\\
\cdots\times\frac{1}{(n_r-1)!}
\left(u_r^{[\nu_1}v_{r[\zeta_1}\delta^{\nu_2\dots\nu_{n_r}]}_{\zeta_2\dots\zeta_{n_r}]}+
v_r^{[\nu_1}u_{r[\zeta_1}\delta^{\nu_2\dots\nu_{n_r}]}_{\zeta_2\dots\zeta_{n_r}]}-
\frac{u_r^{\sigma}v_{r\sigma}}{n_r}\delta^{\nu_1\dots\nu_{n_r}}_{\zeta_1\dots\zeta_{n_r}}\right).
\nonumber
\eea
The result is 
$$
Q^{\alpha}\,{}{}^{\sigma_1\dots\sigma_{s}}_{\gamma_1\dots\gamma_{s}}
{}^{\rho_1\dots\rho_{n_2}\dots\tau_1\dots\tau_{n_r}}_{\epsilon_1\dots\epsilon_{n_2}\dots\zeta_1\dots\zeta_{n_r}}\nabla_{\alpha}
\tilde{A}_{\sigma_1\dots\sigma_s\rho_1\dots\rho_{n_2}\dots\tau_1\dots\tau_{n_r}}=
{\cal J}_{\gamma_1\dots\gamma_{s}\epsilon_1\dots\epsilon_{n_2}\dots\zeta_1\dots\zeta_{n_r}}
$$
where ${\cal J}_{\gamma_1\dots\gamma_{s}\epsilon_1\dots\epsilon_{n_2}\dots\zeta_1\dots\zeta_{n_r}}$ is the contraction of (\ref{hyp}) with the righthand side of (\ref{para2}). This form is manifestly symmetric hyperbolic. Taking into account that the part of $\bm{Q}$ which depends on $\vec{u}_2,\vec{v}_2,\dots,\vec{u}_r,\vec{v}_r$ is non-degenerate when acting on $(r-1)$-fold $(n_2,\dots,n_r)$-forms, this system can in fact be written in the simple form
\bea
(s+1)v^{\rho}\nabla_{[\rho}\tilde{A}_{\gamma_1\dots\gamma_{s}]\epsilon_1\dots\epsilon_{n_2}\dots\zeta_1\dots\zeta_{n_r}}+(-1)^s s v_{[\gamma_1}\nabla^{\rho}
\tilde{A}_{\gamma_2\dots\gamma_{s}]\rho\epsilon_1\dots\epsilon_{n_2}\dots\zeta_1\dots\zeta_{n_r}}=
\nonumber\\
(s+1)
v^{\rho}J_{\rho\gamma_1\dots\gamma_{s}\epsilon_1\dots\epsilon_{n_2}\dots\zeta_1\dots\zeta_{n_r}}
+s v_{[\gamma_1}j_{\gamma_2\dots\gamma_{s}]
\epsilon_1\dots\epsilon_{n_2}\dots\zeta_1\dots\zeta_{n_r}}\, .
\label{newsys}
\eea

Let us finally remark that the hyperbolicity of the general system (\ref{paradigm}) is related to the existence of a wave equation for $A_{\mu_1\dots\mu_m}$. This is immediate from the definition \cite{ES} of the de Rham operator $\Delta_{(1)}$ acting on the first block, which is given ---using the notation of footnote \ref{foot}--- by $\Delta_{(1)}=d_{(1)}\delta_{(1)}+\delta_{(1)}d_{(1)}$. Thus, from (\ref{paradigm}) one deduces
$$
\Delta_{(1)}{A}=-d_{(1)}{j}+(s+1)\delta_{(1)}{J}
$$
which reads, in index notation, \cite{ES}
\bean
\nabla^{\rho}\nabla_{\rho}\tilde{A}_{\mu_1\dots\mu_m}-sR_{\rho[\mu_1}\tilde{A}^{\rho}{}_{\mu_2\dots\mu_s]\mu_{s+1}\dots\mu_m}+\frac{s(s-1)}{2}R_{\rho\sigma[\mu_1\mu_2}
\tilde{A}^{\rho\sigma}{}_{\mu_3\dots\mu_s]\mu_{s+1}\dots\mu_m}\\
+s\sum_{i=s+1}^{m}R^{\rho}{}_{\mu_i\sigma[\mu_1}
\tilde{A}^{\sigma}{}_{\mu_2\dots\mu_s]\mu_{s+1}\dots\mu_{i-1}\rho\mu_{i+1}\dots\mu_m}=
s\nabla_{[\mu_1}j_{\mu_2\dots\mu_s]\mu_{s+1}\dots\mu_{m}}+
(s+1)\nabla^{\rho}J_{\rho\mu_1\dots\mu_m}.
\eean

\section{Constraint equations and integrability conditions}
\label{constraint}
In order to prove the equivalence of the hyperbolized system (\ref{newsys}) with the original (\ref{paradigm}), let us do some counting: the number of unknowns is ${\cal N}C_{n,s}$ where ${\cal N}$ stands for $C_{n,n_2}\cdots C_{n,n_r}$ and $C_{n,s}=n!/(n-s)!s!$. The number of equations in (\ref{paradigm}) is ${\cal N}(C_{n,s-1}+C_{n,s+1})$. There are more equations than unknowns, so that the system is overdetermined. However there are constraint equations ---see \cite{G}---which are easily computed. These are given by, for any $(n-1)$-dimensional hypersurface $\S$ with normal 1-form $N_{\mu}$,
\bea
N^{\rho}\left(\nabla^{\sigma}\tilde{A}_{\sigma\rho\gamma_3\dots\gamma_s
\epsilon_1\dots\epsilon_{n_2}\dots\zeta_1\dots\zeta_{n_r}}-j_{\rho\gamma_3\dots\gamma_s\epsilon_1\dots\epsilon_{n_2}\dots\zeta_1\dots\zeta_{n_r}}\right)=0,
\label{lig1}\\
N_{[\sigma}\nabla_{\gamma_0}\tilde{A}_{\gamma_1\dots\gamma_{s}]\epsilon_1\dots\epsilon_{n_2}\dots\zeta_1\dots\zeta_{n_r}}- N_{[\sigma} J_{\gamma_0\gamma_1\dots\gamma_{s}]\epsilon_1\dots\epsilon_{n_2}\dots\zeta_1\dots\zeta_{n_r}}
=0 \, . \label{lig2}
\eea
Notice that the first of these is absent if $s=n_1\leq 1$, and similarly for the second if $s=n_1\geq n-1$. The total number of constraint equations is therefore ${\cal N}(C_{n-1,s-2}+C_{n-1,s+2})$ and given that
$$
C_{n-1,s-2}+C_{n-1,s+2}=C_{n,s-1}+C_{n,s+1}-C_{n,s}
$$
the constraints are complete.\footnote{By the way, the very same $Q^{\alpha}{}^{\sigma_1\dots\sigma_{n_1}\dots\tau_1\dots\tau_{n_r}}_{\rho_1\dots\rho_{n_1}\dots\nu_1\dots\nu_{n_r}}$ as defined in (\ref{Q}) hyperbolizes directly the system (\ref{trivial}). The resulting system is in fact exactly (\ref{newsys}). However, in this case the constraint equations are not (\ref{lig1},\ref{lig2}) but simply $N_{[\sigma}\nabla_{\gamma_0]}\tilde{A}_{\gamma_1\dots\gamma_{s}\epsilon_1\dots\epsilon_{n_2}\dots\zeta_1\dots\zeta_{n_r}}= N_{[\sigma} J_{\gamma_0]\gamma_1\dots\gamma_{s}\epsilon_1\dots\epsilon_{n_2}\dots\zeta_1\dots\zeta_{n_r}}$ and they are again complete.\label{foot2}} This leads to the equivalence of the new system (\ref{newsys}) to the original one (\ref{paradigm}) together with the constraint equations (\ref{lig1},\ref{lig2}) referred to a spacelike hypersurface with timelike normal $N_{\rho}$. Observe that the system is {\em causal} in the sense that every 1-form $u_{\alpha}$ such that $u_{\alpha}Q^{\alpha}(\cdot,\cdot)$ is positive definite is timelike. Hence, one can define a well-posed {\em Cauchy problem} (or initial value formulation) of the system (\ref{paradigm}) by giving initial data, subject to satisfying the constraints, on any spacelike hypersurface.

The integrability of the constraint equations is not ensured, and depends on the Lorentzian manifold and on the righthand sides of (\ref{paradigm}). The integrability conditions are ruled by the squares of the operators $d_{(1)},\delta_{(1)}$ ---see footnote \ref{foot}---, which are in turn governed by the Ricci identity. Using formulas (23) and (24) in \cite{ES} these integrability conditions are
\bean
\nabla_{[\lambda}J_{\mu\mu_1\dots\mu_{s}]\mu_{s+1}\dots\mu_m}=-\frac{1}{2}\sum_{i=s+1}^{m}R^{\rho}{}_{\mu_i[\lambda\mu}\tilde{A}_{\mu_1\dots\mu_s]\mu_{s+1}\dots\mu_{i-1}\rho\mu_{i+1}\dots\mu_m},\\
\nabla^{\rho}j_{\rho\mu_3\dots\mu_m}=\frac{1}{2}\sum_{i=s+1}^m R^{\rho}{}_{\mu_i\lambda\mu}\tilde{A}^{\lambda\mu}{}_{\mu_3\dots\mu_s\mu_{s+1}\dots\mu_{i-1}\rho\mu_{i+1}\dots\mu_m}.
\eean
Therefore, in flat spacetimes these are rather simple (and easily satisfied), but they may set very strong restrictions on the form of ${J}$ and ${j}$ depending on the algebraic properties of the Riemann tensor. The exceptional cases are (i) when $n_1\leq 1$, in which case the second condition is absent; (ii) the case $n_1\geq n-1$, so that the first condition is trivial; and (iii) when the unknown ${A}$ has $r=1$, i.e., it is a single $m$-form and thus $n_1=m$. Then the previous integrability conditions are simply $\nabla_{[\lambda}J_{\mu\mu_1\dots\mu_{m}]}=0$ and $\nabla^{\rho}j_{\rho\mu_3\dots\mu_m}=0$, which can be enforced in any Lorentzian manifold---for instance by setting ${J}=d{K}$ and ${j}=\delta{k}$ for arbitrary $m$-forms ${K,k}$---; see subsection \ref{m-forms}.

These integrability conditions are reminiscent of the well-known Buchdahl compatibility conditions which arise for spin greater than one in four dimensions \cite{Buch,Buch2}, see also the discussions in \cite{PR,BS}. Of course, they may drastically reduce the number of solutions, or even forbid their existence completely. As a matter of fact, the system (\ref{paradigm}) contains the natural generalization of spin-$S$ fields for {\em any} value of $S$ to arbitrary dimension $n$.

\section{Characteristics}
\label{characteristic}
Let us now consider the important question of the {\em characteristics} of the system (\ref{paradigm}). By definition the characteristics are given by the directions $c_{\alpha}$ such that there are non-trivial solutions $A_{[s][n_2]\dots[n_r]}$ of
$$
c_{\alpha}Q^{\alpha}\,{}{}^{\sigma_1\dots\sigma_{s}}_{\gamma_1\dots\gamma_{s}}
{}^{\rho_1\dots\rho_{n_2}\dots\tau_1\dots\tau_{n_r}}_{\epsilon_1\dots\epsilon_{n_2}\dots\zeta_1\dots\zeta_{n_r}}\,
\tilde{A}_{\sigma_1\dots\sigma_s\rho_1\dots\rho_{n_2}\dots\tau_1\dots\tau_{n_r}}=0.
$$
This condition can be proven to be  equivalent to (setting $\{\Omega\}=\epsilon_1\dots\epsilon_{n_2}\dots\dots\zeta_1\dots\zeta_{n_r}$ to alleviate the notation)
\be
sv_{\rho}c_{[\gamma_1}\tilde{A}^{\rho}{}_{\gamma_2\dots\gamma_s]\{\Omega\}}+
s c_{\rho}v_{[\gamma_1}\tilde{A}^{\rho}{}_{\gamma_2\dots\gamma_s]\{\Omega\}}-
v^{\rho}c_{\rho}\, \tilde{A}_{\gamma_1\dots\gamma_s\{\Omega\}}=0\label{charac}
\ee
whose more general solution is given by either
\begin{enumerate}
\item the null 1-forms $c_{\rho}$ such that $c^{\rho}\tilde{A}_{\rho\gamma_2\dots\gamma_s\{\Omega\}}=0$ and $c_{[\gamma_0}\tilde{A}_{\gamma_1\dots\gamma_s]\{\Omega\}}=0$, so that the solutions have the form $\tilde{A}_{\gamma_1\dots\gamma_s\{\Omega\}}=
c_{[\gamma_1}x_{\gamma_2\dots\gamma_s]\{\Omega\}}$ with $c^{\rho}x_{\rho\dots\gamma_s\{\Omega\}}=0$. \label{i}
\item the spacelike 1-forms $c_{\rho}$ such that $v^{\rho}c_{\rho}=0$ so that the solutions take the form $\tilde{A}_{\gamma_1\dots\gamma_s\{\Omega\}}=
v_{[\gamma_1}c_{\gamma_2}y_{\gamma_3\dots\gamma_s]\{\Omega\}}$ where, without loss of generality, $y_{\gamma_3\dots\gamma_s\{\Omega\}}$ can be taken orthogonal to both $\vec c$ and $\vec v$ in the first block. \label{ii}
\end{enumerate}
As a first conclusion, the characteristics are all {\em physical}, in the sense that they define null, or timelike, hypersurfaces of propagation. A more interesting conclusion is that these characteristics can be characterized, in all cases, by well-defined null directions associated with the solutions. In order to prove this, let us define for the case \ref{ii} two null directions $\{\ell^+_{\mu},\ell^-_\mu\}$ such that 
Span$\{\vec c,\vec v\}=$ Span$\{\vec\ell^+,\vec\ell^-\}$ and $\ell^{+\mu}\ell_{-\mu}=-1$. Then 
$$
\ell^+_{\rho}\ell^+_{[\gamma_1}\tilde{A}^{\rho}{}_{\gamma_2\dots\gamma_s]\{\Omega\}}=0,\hspace{1cm}
\ell^-_{\rho}\ell^-_{[\gamma_1}\tilde{A}^{\rho}{}_{\gamma_2\dots\gamma_s]\{\Omega\}}=0,
$$
which implies that both $\{\ell^+_{\mu},\ell^-_\mu\}$ are in particular {\em principal null directions} (see e.g. Definition 2 in \cite{PP}) of the tensor $A_{[s][n_2]\dots[n_r]}$. The same can be said of the null $\vec c$ in the first case \ref{i} above. It is known \cite{PP} that $\vec k$ is a principal null direction of $A_{[s][n_2]\dots[n_r]}$ if and only if it defines a {\em principal direction} ---see Definition A.2 in \cite{GS}--- of its basic superenergy tensor $T_{\lambda_1\mu_1\dots\lambda_r\mu_r}\{A\}$; that is to say, if and only if 
$k^{\lambda_1}k^{\mu_1}\cdots k^{\lambda_r}k^{\mu_r}T_{\lambda_1\mu_1\dots\lambda_r\mu_r}\{A\}=0$. Due to the general properties of causal tensors  \cite{BS1,GS}, these principal directions are necessarily null. Furthermore, in the case under consideration one is dealing with a special type of principal directions, because they are related to the first block exclusively. This implies that the characteristics of the system (\ref{paradigm}) are defined by the ---necessarily null--- $k^{\rho}$ such that
$$
k^{\lambda_1}k^{\mu_1}T_{\lambda_1\mu_1\dots\lambda_r\mu_r}\{A\}=0
$$
where $k^{\rho}$ stands for either $c^{\rho}$ in case \ref{i} or $\ell^{\pm\rho}$ in case \ref{ii}. 

In summary, {\em all characteristics of the system (\ref{paradigm}) are physical and they are directly related to special principal null directions of the corresponding solutions.}

\section{General energy estimates and inequalities}
\label{inequality}
The hyperbolizations defined by (\ref{Q}) depend on $2r-1$ {\em arbitrary} timelike future-directed vectors (thus, on one timelike vector for $m$-forms, on three timelike vectors for double forms such as the Riemann tensor, and so on). Selecting all of these vectors identical and thus equal to a unique (unit by convenience) timelike vector, say $\vec u$, then $u_{\alpha}Q^{\alpha}(A,A)$ is exactly the ``superenergy" density\footnote{This name may be inappropriate sometimes, because this is the traditional {\em energy} in relevant cases such as the electromagnetic 2-form, or in the case of the massless scalar field $\phi$ by using $d\phi$ as the tensor $A$. However, is some other cases the relevant quantity has been traditionally called ``superenergy", starting with the Bel-Robinson definition for the Weyl tensor \cite{Bel} or the Bel one for the Riemann tensor \cite{Bel1,BoS1}, and also especially for higher order superenergies such as the one defined by using the covariant derivative of the electromagnetic 2-form, see \cite{Ch,S}, or the second derivative of the scalar field, see \cite{S}, references therein, and subsection \ref{Bianchi}.} of the tensor $A_{\mu_1\dots\mu_m}$. This is denoted by $W_{A}(\vec u)$, it can be defined alternatively as
$$
W_A\left(\vec{u}\right)\equiv
T_{\lambda_1\mu_1\dots\lambda_r\mu_r}\{A\}
u^{\lambda_1}u^{\mu_1}\dots u^{\lambda_r}u^{\mu_r} 
$$
or as
$$
W_{A}\left(\vec{u}\right)=\frac{1}{2}\!
\left(\prod_{\Upsilon=1}^{r}\frac{1}{n_{\Upsilon}!}\right)
\tilde{A}_{\mu_1\dots\mu_{n_1},\dots \rho_1\dots\rho_{n_r}}
\tilde{A}_{\nu_1\dots\nu_{n_1}\dots \sigma_1\dots\sigma_{n_r}}
h^{\mu_1\nu_1}\dots h^{\mu_{n_1}\nu_{n_1}}\dots
h^{\rho_1\sigma_1}\dots h^{\rho_{n_r}\sigma_{n_r}}
$$
where
$$
h_{\mu\nu}\left(\vec{u}\right)\equiv g_{\mu\nu}+2u_{\mu}u_{\nu},
$$
and it is equal to half the sum of the positive squares of all the electric-magnetic parts of the tensor $A_{\mu_1\dots\mu_m}$ \cite{S}. The supernergy density is also equal to half the sum of the squares of all the {\em independent} components of $A_{\mu_1\dots\mu_m}$ in any orthonormal basis $\{\vec{e}_{\mu}\}$ with $\vec{e}_0=\vec{u}$ (this is the typical mathematical `energy' of the tensor $A_{\mu_1\dots\mu_m}$):
$$
W_A(\vec{e}_0)=T_{00\dots 0}\{A\}=\frac{1}{2}\left(\prod_{\Upsilon=1}^{r}\frac{1}{n_{\Upsilon}!}\right)
\sum_{\mu_1,\dots,\mu_m=0}^{n-1}
|A_{\mu_1\dots\mu_m}|^2 \, .
$$
This function allows us to obtain estimates because the dominant property implies \cite{S} that $W_A(\vec u)=T_{00\dots 0}\{A\}\geq |T_{\mu_1\dots\mu_m}|$ in any given orthonormal basis. One can also obtain integral (in)equalities related to the system (\ref{paradigm}). As a matter of fact, this quantitiy and the properties of the full superenergy tensors were already used to prove the causal propagation of the gravitational field in \cite{BoS}, and of general physical fields in arbitrary dimension in \cite{BS}. (The causal propagation along characterisitics can also be deduced from the study of the propagation of discontinuities; see  \cite{L,S0}, section 7.3 in \cite{S}, and references therein.) Of course, the Bel-Robinson and Bel tensors {\em in four dimensions} \cite{Bel,Bel1,BoS1} have been repeatedly used to obtain estimates \cite{KN}, and to prove the hyperbolicity of Einstein-Bianchi equations \cite{F,F1,AChY,ChY1,ChY3,ChY2} and the non-linear stability of flat spacetime \cite{CK}. 

Generic integral inequalities are derived by first defining \cite{BS,BoS}
\bean
w(t)\equiv \int_{\overline{D^+(\S_{t_0})}\cap J^-(\S_t)} T_{\lambda_1\mu_1\dots\lambda_r\mu_r}\{A\}
N^{\lambda_1}N^{\mu_1}\dots N^{\lambda_r}N^{\mu_r}  \, \bm\eta =\\
\int_{t_0}^t\left(\int_{\S_{t'}}T_{\lambda_1\mu_1\dots\lambda_r\mu_r}\{A\}
N^{\mu_1}\dots N^{\lambda_r}N^{\mu_r}d\sigma^{\lambda_1}|_{\S_{t'}} \right) dt' \geq 0
\eean
where $\S_{t_0}$ is a compact achronal set (usually a ---piece of--- a hypersurface), $D^+(\S_{t_0})$ its future domain of dependence and $\bm\eta$ denotes the volume element $n$-form. Due to its global hyperbolicity, $D^+(\S_{t_0})$ is foliated by spacelike hypersurfaces $\Sigma_t=\{t=\mbox{constant}\}$ with future normal $N=-dt$ whose  volume element $(n-1)$-form pointing along $\vec N$ is denoted by $d\sigma^{\lambda}|_{\S_{t}}$. Observe that $w(t)=0$ only if $T_{\lambda_1\mu_1\dots\lambda_r\mu_r}\{A\}=0$, which is equivalent to $A_{\mu_1\dots\mu_m}=0$, on $D^+(\S_{t_0})\cap J^-(\S_t)$.  It is very simple to see that $dw(t)/dt\geq 0$
and moreover, using the Gauss theorem one derives for some constant $M$ \cite{BS}
\bea
\frac{dw(t)}{dt}\leq Mw(t) +\int_{\S_{t_0}}T_{\lambda_1\mu_1\dots\lambda_r\mu_r}\{A\}
N^{\mu_1}\dots N^{\lambda_r}N^{\mu_r}d\sigma^{\lambda_1}|_{\S_{t_0}}\nonumber \\
+\int_{\overline{J^-(\S_t)\cap D^+(\S_{t_0})}}\nabla^{\rho}T_{\rho\mu_1\dots\lambda_r\mu_r}\{A\}
N^{\mu_1}\dots N^{\lambda_r}N^{\mu_r}\bm{\eta}\label{ode}
\eea
so that the divergence of the superenergy tensor controls the growth of $w(t)$. But this divergence can be obtained easily using (\ref{set}) and (\ref{E})
\bean
\nabla_{\rho}T^{\rho}{}_{\mu_1\lambda_2\mu_2\dots\lambda_r\mu_r}\{A\}=\frac{1}{2}
E_{\lambda_1\mu_1\dots\lambda_r\mu_r}
{}^{\sigma_1\dots\sigma_{n_1}\dots\tau_1\dots\tau_{n_r}}_{\rho_1\dots\rho_{n_1}\dots \nu_1\dots\nu_{n_r}}\times\hspace{2cm}\\
\left[j_{\beta_2\dots\beta_{n_1}\sigma_1\dots\sigma_{n_2}\dots\tau_1\dots\tau_{n_r}} \tilde{A}_{\mu_1}{}^{\beta_2\dots\beta_{n_1}\rho_1\dots\rho_{n_2}\dots\nu_1\dots\nu_{n_r}}  \right.\hspace{1.5cm}\\
+j^{\beta_2\dots\beta_{n_1}\rho_1\dots\rho_{n_2}\dots\nu_1\dots\nu_{n_r}}
\tilde{A}_{\mu_1\beta_2\dots\beta_{n_1}\sigma_1\dots\sigma_{n_2}\dots\tau_1\dots\tau_{n_r}}\hspace{1cm}\\
-\frac{n_1+1}{n_1}\left(J_{\mu_1\beta_1\dots\beta_{n_1}\sigma_1\dots\sigma_{n_2}\dots\tau_1\dots\tau_{n_r}}\tilde{A}^{\beta_1\dots\beta_{n_1}\rho_1\dots\rho_{n_2}\dots\nu_1\dots\nu_{n_r}}\right.\\
\left.\left.
+J_{\mu_1}{}^{\beta_1\dots\beta_{n_1}\rho_1\dots\rho_{n_2}\dots\nu_1\dots\nu_{n_r}}
\tilde{A}_{\beta_1\dots\beta_{n_1}\sigma_1\dots\sigma_{n_2}\dots\tau_1\dots\tau_{n_r}} \right)\right].
\eean
Notice that a sufficient condition for the superenergy tensor to be divergence-free is that ${j}=0$ and ${J}=0$, see also \cite{PP}. In this particular case, for instance, as well as in the variety of cases in which the last integral in (\ref{ode}) vanishes, (\ref{ode}) can be easily resolved by using the Gronwall Lemma so that
$$
w(t)\leq \frac{1}{M}\left(e^{Mt}-1 \right)\int_{\S_{t_0}}T_{\lambda_1\mu_1\dots\lambda_r\mu_r}\{A\}
N^{\mu_1}\dots N^{\lambda_r}N^{\mu_r}d\sigma^{\lambda_1}|_{\S_{t_0}}\, .
$$
Similar arguments can be used whenever ${j}$ and ${J}$ are polinomic on the tensor $A_{\mu_1\dots\mu_m}$, or even in more general cases.

Observe finally that $T_{\lambda_1\mu_1\dots\lambda_r\mu_r}\{A\}N^{\mu_1}\dots N^{\lambda_r}N^{\mu_r}$ is conserved ---that is to say, it is divergence-free--- if $\nabla_{\rho}T^{\rho}{}_{\mu_1\lambda_2\mu_2\dots\lambda_r\mu_r}\{A\}=0$ and $\vec N$ is a Killing vector \cite{S}.

\section{Examples and applications}
\label{apps}
In this section some selected applications and relevant cases of the general system (\ref{paradigm}) are presented. 
\subsection{$m$-forms}
\label{m-forms}
The simplest application of the results is to the case with $r=1$. The equations are simply
$$
d{A}=(m+1){J}, \hspace{1cm} \delta{A}=-{j}
$$
which are Maxwell-like equations for ordinary $m$-forms $A_{\mu_1\dots\mu_m}=A_{[\mu_1\dots\mu_m]}$. These are of interest in higher dimensional theories such as string theory, supergravity, etcetera. Of course, they include the electromagnetic field equations in any curved background of any dimension by setting $m=2$, ${J}=0$ and letting ${j}$ to be independent of the $2$-form ${A}$. 

From the previous equations one also gets a wave equation 
$$
\Delta{A}=-d{j}+(m+1)\delta{J}
$$
where $\Delta$ is the de Rham operator.

As remarked at the end of section \ref{constraint} the integrability conditions are satisfied whenever $d{J}=0$ and $\delta{j}=0$. The symmetric hyperbolic systems (\ref{newsys}) read in this case
$$
(m+1)v^{\rho}\nabla_{[\rho}A_{\gamma_1\dots\gamma_{m}]}+(-1)^m m\, v_{[\gamma_1}\nabla^{\rho}
A_{\gamma_2\dots\gamma_{m}]\rho}=(m+1)
v^{\rho}J_{\rho\gamma_1\dots\gamma_{m}}
+m\, v_{[\gamma_1}j_{\gamma_2\dots\gamma_{m}]}
$$
for any timelike vector field $\vec v$, which in fact provide all possible hyperbolizations. They are determined by 
$$
T_{\lambda\mu}\{A,B\}=\frac{1}{(m-1)!}\left(
A_{(\lambda}{}^{\rho_2\dots\rho_m}B_{\mu)\rho_2\dots\rho_m}-\frac{1}{2m} A^{\rho_1\rho_2\dots\rho_m}B_{\rho_1\rho_2\dots\rho_m}\right)
$$
which defines the standard energy-momentum tensor of ${A}$ by $T_{\lambda\mu}\{A\}=T_{\lambda\mu}\{A,A\}$. 

Observe that the characteristics of this system are determined by the null eigenvectors of $T_{\lambda\mu}\{A\}$ \cite{BS1,GS}.

\subsection{Extra algebraic relations}
\label{alge}
For the general case of $r$-fold forms with $r>1$, apart from the antisymmetries of each of the $r$ blocks one may have to deal with some additional symmetries involving indices of different blocks. A typical example is that of the Riemann tensor, which is a double (2,2)-form satisfying the first Bianchi identity $R_{\alpha[\beta\lambda\mu]}=0$ which implies the symmetry between the two pairs of indices. Trace-free properties are also in this category. Another typical possibility, in the case of having several fields, is an algebraic relation between them. An example of this is the second Bianchi identity with sources for the gravitational field; see below in subsection \ref{Bianchi}.

The optimal way to deal with these algebraic properties is to simply {\em ignore them} on the unknowns, so that the tensor ${A}$ is not assumed to satisfy them, and {\em add} them as {\em initial conditions} of the Cauchy problem defined by the system of equations. If these properties are preserved by the evolution, then the corresponding solutions will satisfy them.

The strategy, schematically, is as follows. Suppose that ${A}$ is an $r$-fold $(n_1,n_2,\dots,n_r)$-form, and assume that there is an extra symmetry property mixing (say) the first and second block, such as 
$$
\tilde{A}_{[\mu_1\dots\mu_{n_1}\nu_1\dots\nu_q]\nu_{q+1}\dots\nu_{n_2}\dots\dots\tau_1\dots\tau_{n_r}}=0.
$$
Define the $r$-fold $(n_1+q,n_2-q,\dots,n_r)$-form
$$
P_{\mu_1\dots\mu_{n_1+q}\nu_1\dots\nu_{n_2-q}\dots\tau_1\dots\tau_{n_r}}\equiv
\tilde{A}_{[\mu_1\dots\mu_{n_1}\mu_{n_1+1}\dots\mu_{n_1+q}]\nu_{1}\dots\nu_{n_2-q}\dots\tau_1\dots\tau_{n_r}}
$$
and derive differential equations for this tensor field from the original system (\ref{paradigm}). For example, 
$$
\nabla_{[\mu_0}P_{\mu_1\dots\mu_{n_1+q}]\nu_1\dots\nu_{n_2-q}\dots\tau_1\dots\tau_{n_r}}=
J_{[\mu_0\mu_1\dots\mu_{n_1}\mu_{n_1+1}\dots\mu_{n_1+q}]\nu_{1}\dots\nu_{n_2-q}\dots\tau_1\dots\tau_{n_r}}
$$
or
\bean
\nabla^{\rho}P_{\rho\mu_2\dots\mu_{n_1+q}\nu_1\dots\nu_{n_2-q}\dots\tau_1\dots\tau_{n_r}}=
C_{n_1+q-1,q} \,\, j_{\mu_2\dots\mu_{n_1+q}\nu_1\dots\nu_{n_2-q}\dots\tau_1\dots\tau_{n_r}}\\
+(-1)^{n_1+q}(n_1+1)C_{n_1+q-1,q-1}\,\, 
J^{\rho}{}_{[\mu_1\dots\mu_{n_1+q}]\rho\nu_2\dots\nu_{n_2-q}\dots\tau_1\dots\tau_{n_r}}\\
-(-1)^{n_1+q}n_1 C_{n_1+q-1,q-1}\,\, \nabla_{[\mu_1}
\tilde{A}^{\rho}{}_{\mu_2\dots\mu_{n_1+q}]\rho\nu_2\dots\nu_{n_2-q}\dots\tau_1\dots\tau_{n_r}}\, .
\eean
The goal would be to get an homogeneous system whose unique solution, given initial conditions such as the vanishing of ${P}$, is the zero solution. In this particular example a necessary condition is, for instance,  $J_{[\mu_0\mu_1\dots\mu_{n_1}\mu_{n_1+1}\dots\mu_{n_1+q}]\nu_{1}\dots\nu_{n_2-q}\dots\tau_1\dots\tau_{n_r}}=0$.

\subsection{Mixed, or interacting, systems}
\label{mixed}
As explained in the Introduction, one can deal with as many fields satisfying (\ref{paradigm}) as desired, and the righthand sides may then depend on all of them. For example, consider the system (\ref{paradigm}) together with
\be
\nabla_{[\mu_0}\hat{A}_{\mu_1\dots\mu_q]\mu_{q+1}\dots \mu_p}=\hat{J}_{\mu_0\dots \mu_p}, \hspace{1cm}
\nabla^{\rho}\hat{A}_{[\rho\mu_2\dots\mu_q]\mu_{q+1}\dots \mu_p}=\hat{j}_{\mu_2\dots \mu_p}
\label{second}
\ee
where now $J_{\mu_0\dots \mu_m},j_{\mu_2\dots \mu_m},\hat{J}_{\mu_0\dots \mu_p},\hat{j}_{\mu_2\dots \mu_p}$ may in fact depend on both tensor fields $A_{\mu_1\dots \mu_m}$ and $\hat{A}_{\mu_1\dots\mu_p}$. Of course, this can be done for as many fields as desired. 

The hyperbolization of this mixed system is achieved by just hyperbolizing each of the systems separately as shown in section \ref{hyper}, so that the relevant tensors are the corresponding superenergy tensors of ${A}$ and ${\hat{A}}$. If $\hat{r}$ is the form structure number of ${\hat{A}}$, then they depend in general on $2(r+\hat{r}-1)$ arbirary timelike vectors. The total superenergy density ---and hence the total function $w(t)$--- is just the sum of the respective superenergy densities of the tensor fields ${A}$ and ${\hat{A}}$. Note, however, that the corresponding superenergy tensors have different ranks, $2r$ and $2\hat{r}$ respectively, if $r\neq \hat{r}$. Thus, in these cases one simply cannot add them tensorially. Nevertheless, when $r=\hat{r}$ this can be done and in fact this happens in many cases of physical relevance; see subsection \ref{Bianchi}.

In the case of a mixed system (\ref{paradigm})-(\ref{second}), there may be some algebraic relations between the different fields, such as
$$
F_{L}(A,\hat{A},x)=0, \hspace{1cm} L=1,\dots, l\, .
$$
These relations should be treated as extra algebraic relations in the manner explained in subsection \ref{alge}, and thus assumed as initial conditions, checking their preservation by the evolution of the system. 

\subsection{Rank-1 blocks}
\label{rank1}
Now, let us consider the important case in which the tensor field $A_{\mu_1\dots\mu_m}$ has one index with no antisymmetry properties with any other index. In other words, there is a block, say the first, with $n_1=1$. This always includes 1-forms and fully symmetric tensors. It will also be useful in the study of `higher order' systems, see subsection \ref{higher}. Particularizing (\ref{paradigm}) to this case the system reads
\be
\nabla_{[\mu_0}A_{\mu_1]\mu_2\dots\mu_m}=J_{\mu_0\mu_1\dots\mu_m},\hspace{1cm}
\nabla^{\rho}A_{\rho\mu_2\dots\mu_m}=j_{\mu_2\dots\mu_m}\label{sys1}
\ee
where $J_{\mu_0\mu_1\dots\mu_m}=J_{[\mu_0\mu_1]\dots\mu_m}$. This system is always hyperbolizable as 
$$
2v^{\rho}\nabla_{[\rho}A_{\mu_1]\mu_2\dots\mu_m}-v_{\mu_1}\nabla^{\rho}A_{\rho\mu_2\dots\mu_m}=
2v^{\rho}J_{\rho\mu_1\dots\mu_m}
+ v_{\mu_1}j_{\mu_2\dots\mu_m}
$$
for any timelike vector $\vec v$ and this is equivalent to (\ref{sys1}) if the constraint equations
$$
N_{[\sigma}\nabla_{\mu_0}A_{\mu_1]\mu_2\dots\mu_m}- 
N_{[\sigma} J_{\mu_0\mu_1]\mu_2\dots\mu_m}=0
$$
are added, where $\vec N$ is the normal to any $(n-1)$-dimensional hypersurface. These constraints are integrable only if
$$
\nabla_{[\lambda}J_{\mu\mu_1]\mu_{2}\dots\mu_m}=-\frac{1}{2}\sum_{i=2}^{m}
R^{\rho}{}_{\mu_i[\lambda\mu}A_{\mu_{1}]\mu_2\dots\mu_{i-1}\rho\mu_{i+1}\dots\mu_m}\, .
$$
Observe that $j_{\mu_2\dots\mu_m}$ is missing in the constraint equations and their integrability conditions, and thus it is not restricted in any way. This has important implications. For instance, suppose that {\em only} the first equation in (\ref{sys1}) is given. One can obtain a symmetric hyperbolic system in this case by simply {\em supplementing} the second equation in (\ref{sys1}) for an {\em arbitrary} tensor field $j_{\mu_2\dots\mu_m}$. In other words, the divergence on the first index of $A_{\mu_1\dots\mu_m}$ can be considered as a {\em gauge}, not appearing in the original equations, and therefore it can be prescribed at will.  The resulting system is always hyperbolizable as shown in section \ref{hyper}, and one can find the solutions to an initial value problem either depending on $j_{\mu_2\dots\mu_m}$, or alternatively, for any given particular explicit $j_{\mu_2\dots\mu_m}$.

This type of gauge equations can be considered in general, not only for rank-1 blocks; but then the supplementary equations are subject to the corresponding integrability conditions. This general case will be commented upon in subsection \ref{gauge}.

\subsection{Higher order systems}
\label{higher}
In this subsection two kinds of `higher-order' systems will be considered. First, a standard and rather obvious construction of a symmetric hyperbolic system for the first derivative of {\em any} tensor field is presented. Then, the question of higher-order partial differential equations is analyzed using the typical procedure to rewrite them as a first-order system.
\subsubsection{The elementary higher-order systems}
Given {\em any} tensor field $H_{\mu_1\dots\mu_{m-1}}$, subject or not to any differential equations, a symmetric hyperbolic system of equations for {\em its covariant derivative} 
\be
\nabla_{\mu_1}H_{\mu_2\dots\mu_m}\equiv  A_{\mu_1\dots\mu_m}\label{der}
\ee
can always be built. If the original $H_{\mu_1\dots\mu_{m-1}}$ is an $r$-fold $(n_1,\dots,n_r)$-form, then $A_{\mu_1\dots\mu_m}$ is an $(r+1)$-fold $(1,n_1,\dots,n_r)$-form. In particular, the rank of its first block is 1 and this block has no (anti)-symmetries in general with any other index. Furthermore, the Ricci identity implies that
\be
\nabla_{[\mu_0}A_{\mu_1]\mu_2\dots\mu_m}=-\frac{1}{2} \sum_{i=2}^{m}R^{\rho}{}_{\mu_i\mu_0\mu_1}H_{\mu_2\dots\mu_{i-1}\rho\mu_{i+1}\dots\mu_m}\, .
\label{ricci}
\ee
Therefore, the situation is that described in the previous subsection \ref{rank1}. By adding the gauge equations
\be
\nabla^{\rho}A_{\rho\mu_2\dots\mu_m}=j_{\mu_2\dots\mu_m} \label{gau}
\ee
for arbitrarily chosen $j_{\mu_2\dots\mu_m}$---depending on the background, ${A}$ and ${H}$---, one obtains a first-order {\em mixed} symmetric hyperbolic system constituted by the equations (\ref{der},\ref{ricci},\ref{gau}). I will call this system the ``basic higher-order system" for ${H}$. (Observe, by the way, that the last equation (\ref{gau}) is equivalent to a wave equation $
\nabla^{\rho}\nabla_{\rho}H_{\mu_2\dots\mu_m}=j_{\mu_2\dots\mu_m}$).
The basic higher-order system is mixed in the sense of subsection \ref{mixed} and the corresponding parts are given by (i) Eq.(\ref{der}) by itself and (ii) the two equations (\ref{ricci},\ref{gau}) for $A_{\mu_1\dots\mu_m}$. The part (i) is of type (\ref{trivial}) and thus hyperbolized in the way outlined in footnote \ref{foot2}. Besides, its integrability conditions are {\em always} satisfied by virtue of (\ref{ricci}). The second part (ii) is of the generic type (\ref{paradigm}) treated in this paper, and belongs to the particular case discussed in the previous subsection. Its integrability conditions are {\em also} satisfied by virtue of the second Bianchi identity for the Riemann tensor and (\ref{der}) itself. 

A more drastic way of building a symmetric hyperbolic system is to give the whole symmetric derivative of $A_{\mu_1\dots\mu_m}$, and not only its trace. Thus, instead of (\ref{gau}) one can add
\be
\nabla_{(\mu_0}A_{\mu_1)\mu_2\dots\mu_m}=\chi_{\mu_0\mu_1\mu_2\dots\mu_m}\label{gau2}
\ee
for arbitrary $\chi_{\mu_0\mu_1\mu_2\dots\mu_m}=\chi_{(\mu_0\mu_1)\mu_2\dots\mu_m}$. This leads, together with (\ref{ricci}), to a system of type (\ref{trivial}).

Of course, if $H_{\mu_1\dots\mu_{m-1}}$ was subject to some differential equations, then the choice of $j_{\mu_2\dots\mu_m}$ in (\ref{gau}) or $\chi_{\mu_0\mu_1\mu_2\dots\mu_m}$ in (\ref{gau2}) must be done accordingly, in a compatible manner. From the perspective of the basic higher-order system, any differential equations on ${H}$ can be seen as extra algebraic relations in the sense of subsection \ref{alge}, and they simply read
$$
F_{L}(A,H,x)=0, \hspace{1cm} L=1,\dots, l
$$
for some functions $F_L$.
Thus, as a first criterion for the solvability of any equations on $H_{\mu_1\dots\mu_{m-1}}$, one can always construct its basic higher-order system, hyperbolize it, and check if the original equations, considered now as initial conditions of the previous type, are preserved by the evolution of the system.

\subsubsection{Higher order partial differential equations}
Consider now the case of a system of higher-order partial differential equations for an unknown tensor field $L_{\mu_1\dots\mu_m}$. Assume that the order of the system is $k$ so that it can be written as
\be
K_{\Omega}{}^{\rho_1\dots\rho_k}{}_{\mu_1\dots\mu_m}\nabla_{\rho_1}\dots\nabla_{\rho_k}L^{\mu_1\dots\mu_m}=I_{\Omega}\label{k-order}
\ee
for some functions $I_{\Omega},K_{\Omega}{}^{\rho_1\dots\rho_k}{}_{\mu_1\dots\mu_m}$ depending on the background Lorentzian manifold, and on $L_{\mu_1\dots\mu_m}$ and its covariant derivatives up to order $k-1$.   Without loss of generality assume that $K_{\Omega}{}^{(\rho_1\dots\rho_k)}{}_{\mu_1\dots\mu_m}=K_{\Omega}{}^{\rho_1\dots\rho_k}{}_{\mu_1\dots\mu_m}$. By introducing auxiliary fields corresponding to each of the symmetrized derivatives of $L_{\mu_1\dots\mu_m}$ up to order $k-1$ as follows \cite{G}
\bea
\nabla_{\nu_1}L_{\mu_1\dots\mu_m}&\equiv& A_{\nu_1\mu_1\dots\mu_m}, \label{cero}\\
\nabla_{(\nu_1}A_{\nu_2\dots\nu_i)\mu_1\dots\mu_m}&\equiv& A_{\nu_1\dots\nu_i\mu_1\dots\mu_m} , \hspace{0.3cm} i=2,\dots, k-1 \nonumber
\eea
one can then construct $k-2$ basic higher-order systems in the sense above by using the corresponding Ricci identities when needed:
\bea
\nabla_{\nu_2}A_{\nu_1\mu_1\dots\mu_m}&=&A_{\nu_1\nu_2\mu_1\dots\mu_m}+
\frac{1}{2}\sum_{j=1}^m R^{\rho}{}_{\mu_j\nu_2\nu_1}L_{\mu_1\dots\mu_{j-1}\rho\mu_{j+1}\dots\mu_m}, \label{uno}\\
\nabla_{\nu_3}A_{\nu_2\nu_1\mu_1\dots\mu_m}&=&A_{\nu_1\nu_2\nu_3\mu_1\dots\mu_m}+
\sum_{j=1}^m
R^{\rho}{}_{\mu_j\nu_3(\nu_2}A_{\nu_1)\mu_1\dots\mu_m}-
\frac{2}{3}R^{\rho}{}_{(\nu_1\nu_2)\nu_3}A_{\rho\mu_1\dots\mu_m}\nonumber\\
&&+\frac{1}{3}\sum_{j=1}^m\left(\nabla_{\nu_1}R^{\rho}{}_{\mu_j\nu_3\nu_2}+
\nabla_{\nu_2}R^{\rho}{}_{\mu_j\nu_3\nu_1}\right)L_{\mu_1\dots\mu_{j-1}\rho\mu_{j+1}\dots\mu_m}
\label{dos}\\
\vdots\hspace{2cm} \vdots &&\vdots\hspace{2cm}\vdots\hspace{2cm}\vdots\hspace{2cm}\vdots\nonumber
\eea
and so on for all $i=1,\dots, k-2$, together with the last equation derivable from a Ricci identity, namely
\be
\nabla_{[\nu_1}A_{\nu_2]\nu_3\dots\nu_k\mu_1\dots\mu_m}={\cal F}_{\nu_1\nu_2\nu_3\dots\nu_k\mu_1\dots\mu_m}\label{tres}
\ee
where ${\cal F}_{\nu_1\nu_2\nu_3\dots\nu_k\mu_1\dots\mu_m}={\cal F}_{[\nu_1\nu_2](\nu_3\dots\nu_k)\mu_1\dots\mu_m}$ is a known function depending on the background spacetime and the $A_{\nu_1\dots\nu_i\mu_1\dots\mu_m}$ for $i=1,\dots ,k-2$. The result is a mixed first-order system formed by (\ref{cero}-\ref{tres}) together with the original higher-order equation (\ref{k-order}) rewritten as 
\be
K_{\Omega}{}^{\rho_1\dots\rho_k}{}_{\mu_1\dots\mu_m}\nabla_{\rho_1}A_{\rho_2\dots\rho_k\mu_1\dots\mu_m}=I_{\Omega}\label{cuatro}
\ee
where $I_{\Omega},K_{\Omega}{}^{\rho_1\dots\rho_k}{}_{\mu_1\dots\mu_m}$ are  now considered as functions of $L_{\mu_1\dots\mu_m}$ and $A_{\nu_1\dots\nu_i\mu_1\dots\mu_m}$ for $i=1,\dots,k-1$. Each of the systems (\ref{cero}), (\ref{uno}), (\ref{dos}), etcetera, are of the type (\ref{trivial}), and thus easily hyperbolized as indicated in the footnote \ref{foot2}. The constraint equations for these systems are complete and integrable by virtue of (\ref{uno},\ref{dos}, ..... ,\ref{tres}), respectively. The remaining system is formed by (\ref{tres}) and (\ref{cuatro}), for which the constraints are complete and integrable too. The question of whether this system is hyperbolizable or not depends on the explicit form of $K_{\Omega}{}^{\rho_1\dots\rho_k}{}_{\mu_1\dots\mu_m}$. Some simple hyperbolizable examples leading to symmetric hyperbolic systems obtained by the procedure explained in section \ref{hyper} are (see also \cite{G})
$$
K_{\Omega}{}^{\rho_1\dots\rho_k}{}_{\mu_1\dots\mu_m}=g^{\rho_1(\rho_2}\delta^{\rho_3}_{\sigma_3}\cdots\delta^{\rho_k)}_{\sigma_k}\Delta_{\mu_1\dots\mu_m}^{\beta_1\dots\beta_m}
$$
or
$$
K_{\Omega}{}^{\rho_1\dots\rho_k}{}_{\mu_1\dots\mu_m}=\delta^{(\rho_1}_{\sigma_1}\cdots\delta^{\rho_k)}_{\sigma_k}\Delta_{\mu_1\dots\mu_m}^{\beta_1\dots\beta_m}
$$
where in both cases $\Delta_{\mu_1\dots\mu_m}^{\beta_1\dots\beta_m}$ represents the product $\delta_{\mu_1}^{\beta_1}\cdots\delta_{\mu_m}^{\beta_m}$ with the appropriate (anti)-symmetrizations on the indices $\mu_1\dots\mu_m$ according to the $r$-fold form structure of $L_{\mu_1\dots\mu_m}$ or the symmetries between its indices.

Of course, a particular explicit very simple example of the above is the massless scalar field $\phi$ which satisfies the second order equation
$$
\nabla^{\rho}\nabla_{\rho} \phi =0 \, .
$$
Then, one can construt a first order system on the variables $(\phi,\phi_{\mu})$ by means of
$$
\nabla_{\mu}\phi \equiv \phi_{\mu}, \,\,\, \nabla_{[\nu}\phi_{\mu]}=0, \,\,\, \nabla^{\rho}\phi_{\rho}=0.
$$
This is the case of a 1-form, $m=1$ in subsection \ref{m-forms}. The hyperbolizations are defined by the energy-momentum tensor of the scalar field
\be
T_{\lambda\mu}\{\nabla\phi\}=
\phi_{\lambda}\phi_{\mu}-\frac{1}{2}g_{\lambda\mu}\phi^{\rho}\phi_{\rho}\, .\label{ses}
\ee
Thus, the characteristics of this system are determined by the null eigenvectors of $T_{\lambda\mu}\{\nabla\phi\}$ \cite{BS1,GS}, and they depend on the causal character of $\phi_{\mu}$.

\subsection{Adding {\em `gauge'} equations}
\label{gauge}
Consider the case when only the first part of the system (\ref{paradigm}) is given, namely
$$
\nabla_{[\mu_0}A_{\mu_1\dots\mu_s]\mu_{s+1}\dots \mu_m}=J_{\mu_0\dots \mu_m}
$$
and assume, without loss of generality, that $s=n_1$ is the first block rank of ${A}$ seen as an $r$-fold form. These equations by themselves are underdetermined if $s>(n-1)/2$, because there are ${\cal N}C_{n,s}$ unknowns and only ${\cal N}C_{n,s+1}$ equations. Therefore, for these values of $s$, there are no possible hyperbolizations of these equations as they stand \cite{G}. 

However, a feasible procedure for {\em general} values of $s$ is the following. Observing that the symmetric part (with respect to an index in the first block) of the covariant derivative of ${A}$ is not restricted in any way by the given equations, one can add these, fully or partly,  {\em arbitrarily} as `gauge' equations. Thus, for instance, one could add either of
$$
\nabla_{(\mu_0}A_{\mu_1)\mu_2\dots\mu_s\mu_{s+1}\dots \mu_m}=
\hat{J}_{\mu_0\dots \mu_m}, \hspace{1cm}
\nabla^{\rho}A_{\rho\mu_2\dots\mu_m}=j_{\mu_2\dots\mu_m}
$$
for arbitrary $\hat{J}_{\mu_0\mu_1\mu_2\dots \mu_m}=\hat{J}_{(\mu_0\mu_1)[\mu_2\dots\mu_s]\mu_{s+1}\dots \mu_m}$ or $j_{\mu_2\dots\mu_s\mu_{s+1}\dots \mu_m}=j_{[\mu_2\dots\mu_s]\mu_{s+1}\dots \mu_m}$. This leads, in the first case, to a system of type (\ref{trivial}), and in the second to a system of type (\ref{paradigm}). Any of these completed systems can be hyperbolized as explained in previous sections. The only restrictions to be placed on $\hat{J}_{\mu_0\mu_1\mu_2\dots \mu_m}$ or $j_{\mu_2\dots\mu_s\mu_{s+1}\dots \mu_m}$ are the necessary integrability conditions written down in section \ref{constraint}. Then, one can find the solutions to an initial value problem depending on the presribed  $j_{\mu_2\dots\mu_m}$ (or $\hat{J}_{\mu_0\mu_1\mu_2\dots \mu_m}$). Alternatively, one can solve the system, depending on initial data, for any given particular explicit $j_{\mu_2\dots\mu_m}$ (or $\hat{J}_{\mu_0\mu_1\mu_2\dots \mu_m}$).

Similarly, if only the second in (\ref{paradigm}) is given, one can add the first in (\ref{paradigm}) as a gauge for arbitrary $J_{\mu_0\dots \mu_m}$, subject to its integrability conditions. 

\subsection{Bianchi identities with and without sources in arbitrary dimension}
\label{Bianchi}
As an important physical application of all of the above, let us consider the so-called {\em higher-order} field equations for the gravitational field, see \cite{L}, which are essentially the Bianchi identities for the Riemann tensor. Letting $K_{\alpha\beta\lambda\mu}=K_{[\alpha\beta][\lambda\mu]}$ be {\em any} double (2,2)-form, {\em not} necessarily satisfying any extra symmetry property such as $K_{[\alpha\beta\lambda]\mu}=0$, the `Bianchi' equations are
\be
\nabla_{[\gamma}K_{\alpha\beta]\lambda\mu}=0. \label{Bi}
\ee
This is an underdetermined system when $n=4$, but {\em not} for higher values of the dimension $n$. In any case, as explained in the previous subsection \ref{gauge}, one can add for arbitrary $n$ a `gauge' part
\be
\nabla^{\rho}K_{\rho\beta\lambda\mu}=j_{\beta\lambda\mu}\label{Bi2}
\ee
where $j_{\rho\lambda\mu}=j_{\rho[\lambda\mu]}$ is restricted to satisfy the integrability conditions
\be
\nabla^{\rho}j_{\rho\lambda\mu}=\frac{1}{2}\left(R^{\rho}{}_{\lambda\tau\nu}K^{\tau\nu}{}_{\rho\mu} + R^{\rho}{}_{\mu\tau\nu}K^{\tau\nu}{}_{\lambda\rho}\right)\label{int}
\ee
but is otherwise arbitrary. 

These equations can be attacked in two completely different ways, both from the mathematical and physical points of view. The first possibility is taking the background Lorentzian manifold as given ---and therefore its Riemann tensor is known data---, and treating $K_{\alpha\beta\lambda\mu}$ as an unknown tensor field to be determined. This has some interest from (i) the mathematical point of view, as one of the solutions is the Riemann tensor itself and (ii) from the physical viewpoint, because it serves the purpose of studying ``spin-2" fields on given spacetimes, such as a linearized field in flat spacetime and similar cases. The second possibility, which is much more difficult to handle and has a richer structure and more physical interest, is the case when the background spacetime is not given, but one wants to find the metric tensor field and corresponding connection alongside the solutions for the higher-order equations (\ref{Bi}). In this case, one has to build a hyperbolic reduction of the Einstein's field equations for some matter content where the unknowns contain metric components, connection variables, and curvature components. The way to do this in four dimensions, which can be translated in principle to higher dimensions, can be consulted in \cite{F,F1,AChY,ChY1,ChY3,ChY2}.

In both cases, the hyperbolization of the Bianchi equations by themselves, or as part of a larger mixed system, should be achieved. Considering the system (\ref{Bi}-\ref{Bi2}), and by the general results of this paper, the constraint equations are complete, a symmetric hyperbolization is provided by the superenergy tensor of $K_{\alpha\beta\lambda\mu}$ (its Bel tensor, see \cite{Bel1,BoS1,S}):
\bea
T_{\alpha\beta\lambda\mu}\left\{K_{[2][2]}\right\}=
K_{\alpha\rho\lambda\sigma}
K_{\beta}{}^{\rho}{}_{\mu}{}^{\sigma}
+K_{\alpha\rho\mu\sigma}
K_{\beta}{}^{\rho}{}_{\lambda}{}^{\sigma}\nonumber\\
-\frac{1}{2}g_{\alpha\beta}
K_{\rho\tau\lambda\sigma}K^{\rho\tau}{}_{\mu}{}^{\sigma}
-\frac{1}{2}g_{\lambda\mu}
K_{\alpha\rho\sigma\tau}K_{\beta}{}^{\rho\sigma\tau}+
\frac{1}{8}g_{\alpha\beta}g_{\lambda\mu}
K_{\rho\tau\sigma\nu}
K^{\rho\tau\sigma\nu} \label{seK}
\eea
and the characerisitics of the system (\ref{Bi}-\ref{Bi2}) are determined by the principal directions of $T_{\alpha\beta\lambda\mu}\left\{K_{[2][2]}\right\}$ such that
$$
\ell^{\alpha}\ell^{\beta}T_{\alpha\beta\lambda\mu}\left\{K_{[2][2]}\right\}=0
$$
which are the principal null directions of the corresponding $K_{[2][2]}$, defined by
$$
\ell^{\rho}\ell_{[\alpha}K_{\beta]\rho\lambda\mu}=0.
$$
Furthermore, the divergence of $T_{\alpha\beta\lambda\mu}\left\{K_{[2][2]}\right\}$ can be easily computed usign (\ref{Bi}-\ref{Bi2}) to give
$$
\nabla^{\alpha}T_{\alpha\beta\lambda\mu}\left\{K_{[2][2]}\right\}=
j_{\rho\lambda\sigma}K_{\beta}{}^{\rho}{}_{\mu}{}^{\sigma}+
j_{\rho\mu\sigma}K_{\beta}{}^{\rho}{}_{\lambda}{}^{\sigma}-\frac{1}{2}
j_{\rho\tau\sigma}K_{\beta}{}^{\rho\tau\sigma}\, .
$$
and this rules, as explained in section \ref{inequality}, the (super)-energy estimates.

All of the above can be said in general. Nevertheless, one can consider the important particular case of Einstein's field equations
$$
R_{\mu\nu}-\frac{1}{2}Rg_{\mu\nu}+\Lambda g_{\mu\nu}=\frac{8\pi G}{c^4}T_{\mu\nu}, \hspace{1cm} R=R^{\rho}{}_{\rho}
$$
or similar problems for most general equations relating the Ricci tensor $R_{\mu\nu}$ and the energy-momentum tensor $T_{\mu\nu}$ of the gravitational sources, which must be divergence-free by assumption
\be
\nabla^{\rho}T_{\rho\mu}=0.\label{div}
\ee
Given that the Riemann tensor satisfies
\be
R_{\alpha[\beta\lambda\mu]}=0 \,\, \Longrightarrow \,\, 
R_{\alpha\beta\lambda\mu}=R_{\lambda\mu\alpha\beta}\label{cyclic}
\ee
from (\ref{Bi}) one immediately deduces 
$$
\nabla^{\rho}R_{\rho\beta\lambda\mu}=2\nabla_{[\lambda}R_{\mu]\beta}
$$
so that an appropriate choice for the arbitrary $j_{\beta\lambda\mu}$ of (\ref{Bi2}) is
$$
j_{\beta\lambda\mu}=\frac{16\pi G}{c^4}\left(\nabla_{[\lambda}T_{\mu]\beta}-
\frac{1}{n-2}g_{\beta[\mu}\nabla_{\lambda]}T \right), \hspace{1cm} T=T^{\rho}{}_{\rho}\, .
$$
This choice is needed if one wishes that the solutions of (\ref{Bi}-\ref{Bi2}) include the Riemann tensor of the Lorentzian metric. The integrability condition (\ref{int}) reduces now, due to (\ref{div}) which implies $\nabla^{\rho}j_{\rho\lambda\mu}=0$, to
\be
R^{\rho}{}_{\lambda\tau\nu}K^{\tau\nu}{}_{\rho\mu} + R^{\rho}{}_{\mu\tau\nu}K^{\tau\nu}{}_{\lambda\rho}=0 \label{c1}
\ee
while the integrability conditions of the original (\ref{Bi}) are
\be
R^{\rho}{}_{\lambda[\tau\nu}K_{\alpha\beta]\rho\mu}+
R^{\rho}{}_{\mu[\tau\nu}K_{\alpha\beta]\lambda\rho}=0\, .\label{c2}
\ee
The extra algebraic relations (\ref{cyclic}) can be treated in the manner explained in section \ref{alge}, by defining the tensor
$$
P_{\alpha\beta\lambda\mu}\equiv K_{[\alpha\beta\lambda]\mu}
$$
and imposing $P_{\alpha\beta\lambda\mu}=0$ as initial conditions. This can be shown to be preserved by the evolution, see \cite{AChY,Bo}. 

Similarly, the Einstein equations themselves can be treated as initial conditions by the method of subsection \ref{mixed}, {\em whenever the explicit form of the energy-momentum tensor in terms of the gravitating physical fields is given}. (This has been done in 4 dimensions for some cases in \cite{ChY3,ChY2}.) I am going to illustrate the method by considering the case in which there is only a massless scalar field $\phi$ on the spacetime.

The energy-momentum tensor of the scalar field is given by (\ref{ses}), hence the needed choice for $j_{\beta\lambda\mu}$  is, using the notation introduced at the end of subsection \ref{higher},
$$
j_{\beta\lambda\mu}=\frac{16\pi G}{c^4}\nabla_{\beta}\phi_{[\lambda}\phi_{\mu]}\, .
$$
This suggests that the proper thing to do is going one step higher for the scalar field defining
$$
\phi_{\mu\nu}=\nabla_{\mu}\phi_{\nu} \, , \hspace{1cm} \phi_{\mu\nu}=\phi_{(\mu\nu)}
$$
whose field equations are easily determined using the method of subsection \ref{higher}:
$$
\nabla_{[\lambda}\phi_{\mu]\nu}=-\frac{1}{2}\phi_{\rho}R^{\rho}{}_{\nu\lambda\mu}\, , \hspace{1cm}
\nabla^{\rho}\phi_{\rho\nu}=-\phi^{\rho}R_{\rho\nu}\, .
$$
The Einstein field equations read simply
$$
R_{\mu\nu}-\frac{8\pi G}{c^4}\phi_{\mu}\phi_{\nu}+\frac{2}{n-2}\Lambda g_{\mu\nu}=0\, .
$$
Therefore, the system to be considered is a mixed system, for the unknowns $(K_{\alpha\beta\lambda\mu},\phi_{\mu\nu},\phi_{\mu})$ where $K_{\alpha\beta\lambda\mu}$ is a double (2,2)-form, $\phi_{\mu\nu}$ is a double symmetric (1,1)-form and $\phi_{\mu}$ is a one-form. The equations are (on a given background)
\bean
\nabla_{\mu}\phi_{\nu} =\phi_{\mu\nu}\, , \hspace{1cm} \nabla_{[\lambda}\phi_{\mu]\nu}=-\frac{1}{2}\phi_{\rho}R^{\rho}{}_{\nu\lambda\mu}\, , \hspace{1cm}
\nabla^{\rho}\phi_{\rho\nu}=\phi_{\nu}\left(\frac{2}{n-2}\Lambda-
\frac{8\pi G}{c^4}\phi^{\rho}\phi_{\rho}\right)\\
\nabla_{[\gamma}K_{\alpha\beta]\lambda\mu}=0\, , \hspace{1cm}
\nabla^{\rho}K_{\rho\beta\lambda\mu}=
\frac{16\pi G}{c^4}\phi_{\beta[\lambda}\phi_{\mu]}\, .\hspace{3cm}
\eean
This must be supplemented with the extra algebraic relations (\ref{c1}), (\ref{c2}) together with
$$
\phi^{\rho}{}_{\rho}=0, \hspace{1cm} K_{[\alpha\beta\lambda]\mu}=0, \hspace{1cm}
K^{\rho}{}_{(\mu\nu)\rho}+\frac{8\pi G}{c^4}\phi_{\mu}\phi_{\nu}+\frac{2}{n-2}\Lambda g_{\mu\nu}=0
$$
all of which can be considered as initial conditions for the data on an initial spacelike hypersurface $\S$. Of course, this data must be restricted to satisfy the constraint equations provided by (\ref{lig1}-\ref{lig2}):
\bean
N_{[\lambda}\left(\nabla_{\mu]}\phi_{\nu} -\phi_{\mu]\nu}\right)=0, \hspace{3mm}
N_{[\tau} \nabla_{\lambda}\phi_{\mu]\nu}+\frac{1}{2}\phi_{\rho}R^{\rho}{}_{\nu[\lambda\mu}N_{\tau]}=0,\\
N_{[\tau} \nabla_{\gamma}K_{\alpha\beta]\lambda\mu}=0, \hspace{3mm} 
N^{\beta}\left(\nabla^{\rho}K_{\rho\beta\lambda\mu}-
\frac{16\pi G}{c^4}\phi_{\beta[\lambda}\phi_{\mu]}\right)=0
\eean
where $\vec N$  is the normal to $\S$.
This system of equations is directly hyperbolizable as shown in previous sections by using superenergy tensors. Here, apart from the Bel tensor (\ref{seK}) mentioned above, one has to use the superenergy tensor of the scalar field (\cite{S} and references therein):
$$
T_{\alpha\beta\lambda\mu}\left\{\phi_{[1][1]}\right\}=
2\phi_{\alpha(\lambda} \phi_{\mu)\beta}
- g_{\alpha\beta}\phi_{\lambda}{}^{\rho}\phi_{\mu\rho}
-g_{\lambda\mu}\phi_{\alpha}{}^{\rho}\phi_{\beta\rho}
+\frac{1}{2} g_{\alpha\beta}g_{\lambda\mu}
\phi^{\sigma\rho}\phi_{\sigma\rho} \, .
$$
Observe that both needed superenergy tensors have the same number of indices and the same symmetry properties. They can in fact be added into one single, mixed, superenergy tensor
$$
T^{(tot)}_{\alpha\beta\lambda\mu}\equiv T_{\alpha\beta\lambda\mu}\left\{K_{[2][2]}\right\}+T_{\alpha\beta\lambda\mu}\left\{\phi_{[1][1]}\right\}
$$
whose total superenergy density relative to a timelike unit vector $\vec u$, and the total $w(t)$ of section \ref{inequality}, is the sum of the separate corresponding quantities for each of the two fields involved. It is very interesting to remark that the inequalities computed in section \ref{inequality} can then be easily estimated by using the divergence of the total tensor above \cite{S}. Furthermore, in cases with Killing vectors, the corresponding {\em total, mixed, superenergy currents} are divergence free, and provide conserved quantities, see subsection 7.2 in \cite{S}.

A similar procedure can be used to couple the Bianchi equations to other matter fields, such as the electromagnetic one. In that case, one must use the symmetrized superenergy tensor associated to the covariant derivative of the Maxwell 2-form, which is the Chevreton tensor \cite{Ch,S}.
As a matter of fact, one can go to higher-order derivatives of all the fields involved (the Riemann tensor, the scalar field, the electromagentic 2-form, etcetera) and find hyperbolizations of the new variables by using the different (super)$^k$-energy tensors, for all natural numbers $k$, introduced in \cite{S}.

\section*{Acknowledgements}
Thanks to S.B. Edgar for reading the manuscript.
Financial support under
grants BFM2000-0018 of the Spanish CICyT and 
no. 9/UPV 00172.310-14456/2002 of the University of the Basque 
Country is acknowledged.

\end{document}